\documentstyle[preprint,aps]{revtex}
\begin{document}
\draft
\preprint{UCF-CM-93-004 (revised)}
\title
{Mesoscopic Transport Beyond Linear Response}
\author{O. Heinonen\cite{byline} and M.D. Johnson}
\address{
Department of Physics, University of Central Florida, Orlando, FL 32816
}
\maketitle
\begin{abstract}
We present an approach to steady-state mesoscopic transport based on the
maximum entropy principle formulation of nonequilibrium statistical mechanics.
Our approach is not limited to the linear response regime.
We show that this approach
yields the quantization observed in the integer quantum
Hall effect at large currents, which until now has been unexplained.
We also predict new behaviors of
non-local resistances at large currents in the
presence of dirty contacts.
\end{abstract}
\pagebreak

\narrowtext

In this Letter, we propose a non-perturbative general approach to nonlinear
nonequilibrium steady-state transport in mesoscopic systems. Our work is
based on the maximum entropy approach (MEA) to nonequilibrium
statistical mechanics \cite{MEP}, in which
the density matrix is found by
by maximizing the information entropy of the system, subject to constraints
which fix the expectation values of observables.
Although the MEA should in principle be
applicable to any nonequilibrium system, examples and explicit
calculations have in practice been limited.
In part this is because it is seldom possible to calculate the microstates of
nonequilibrium systems.
Moreover, it is difficult in general to determine whether the
information entropy is equal to the thermodynamic entropy, as is needed,
{\it e.g.}, to identify the temperature in the density matrix derived from the
MEA.
The problem of steady-state mesoscopic transport, however, is uniquely
well suited to this approach: the microstates can be calculated to obtain
the exact density matrix,
and recent work by Hershfield \cite{Hershfield} allows us to identify
the temperature in the density matrix.
Here we show how the MEA can be used to calculate
nonlinear current-voltage relations in mesoscopic devices.

Our study was initially motivated by a very important but often neglected
fact: the integer quantum Hall effect (IQHE) is exhibited
even in systems driven by very large currents \cite{NIST}.  The IQHE
can be viewed as a near-ideal manifestation of mesoscopic transport
\cite{suppression}.  An elegant explanation of the IQHE at low
currents is given by the
Landauer-B\"uttiker (LB) approach \cite{LB} to mesoscopic transport.
However, this approach is
fundamentally a linear response theory \cite{ADStone,BarangerStone}, and, as
we show below, fails to
yield the quantization observed at high currents \cite{VanSon}.
Hence the IQHE exhibits
a `simple' behavior (perfect quantization) far
beyond the linear response regime.
There exist other, general, approaches to nonequilibrium transport,
such as various Green's function techniques
\cite{KadanoffBaym,Keldysh,Ferry,Datta}. These lead to
quite complicated calculations even in the linear regime, and it
is not clear whether they can yield the IQHE in the nonlinear regime.
Any comprehensive theory
of nonlinear mesoscopic
transport must be able to explain the extraordinary quantization of the
IQHE at high currents.
One of our most important results is to show that this
can in fact be explained within the MEA.

We consider
a mesoscopic system consisting of a device ({\em e.g.\/} a Hall
bar or a quantum wire), to which $M$ terminals, denoted by $m$
($m=s,d,1,2,\ldots,M-2$) are connected by straight leads long enough
that evanescent modes emanating from the terminals
decay to zero in the leads.
A current $I$ flows from the source $s$ to the drain $d$.
Electrons injected in states at one terminal can either be transmitted
to another terminal, or reflected back.  In either case, they lose their
phase memory upon re-entering a terminal due to phase-randomizing
scattering there.
The electrons in the system are described by a complete orthogonal set of
eigenstates $|\psi_\alpha\rangle$ with energies $\epsilon_\alpha$.
Each eigenstate in general carries a net
current $i_{m,\alpha}$ from each terminal\cite{convention}.
A particularly useful set
of eigenstates for multi-terminal ($M>2$) systems
are the scattering states \cite{BarangerStone,Sols}
$|\psi^+_{mnk}\rangle$ used when the terminals are modeled as
semi-infinite straight leads. The state $|\psi^+_{mnk}\rangle$ is
incoming into the device from terminal $m$; $n$ and $k$ denote
the asymptotic wavenumber and subband index of the incoming wave. With a
proper normalization\cite{Sols}, this state's net
current, $i_{m',mnk}$, at terminal $m'$ is related to the incoming
current $i^0_{mnk}$ by
$i_{m',mnk}=i^0_{mnk}(\delta_{m'm}-\sum_{n'k'}
t_{m'n'k',mnk})$, where $t_{m'n'k',mnk}$ is the transition probability
obtained from the scattering matrix in the $|\psi^+_{mnk}\rangle$
representation.

In the LB formalism \cite{LB} it is assumed that each
terminal is held at a `local chemical potential' $\mu_m$, so
that electrons are injected into
the device at each terminal with distributions
$f_{mnk}^{LB} = 1/[e^{\beta(\epsilon_{mnk}-\mu_{m})}+1]$.
For a two-terminal system at zero temperature and low
voltages $V=(\mu_s-\mu_d)/e$, this gives
a resistance $R=h/(je^2\widetilde t\,)$.
Here $j$ is the number of occupied subbands
and $\widetilde t$ is the total transmission probability at
$\mu_s$\cite{spinless}.
This resistance is quantized in the absence of backscattering
($\widetilde t=1$);
this is the two-terminal version of the LB explanation of the low-current IQHE.
However, when the voltage becomes greater than the subband spacing, the source
injects electrons into the $(j+1)$st subband but the
drain does not. Then, according to the LB approach, the resistance of an
ideal system would lie between $h/(je^2)$ and $h/[(j+1)e^2]$. The
same conclusion is reached for the case of a multi-terminal system.
(We point out that this argument has been invoked to explain the large-voltage
failure of resistance quantization in quantum point contact experiments
within the LB formalism \cite{van Houten}.)
Yet in precision IQHE measurements, the voltage is many times greater than
the subband spacing and the resistance is nonetheless highly
quantized \cite{NIST}.
Hartree interactions moves subbands together and cannot
change this conclusion.
Exchange-correlation
interactions could restore quantization if
they caused a large energy separation (of order $eV$)
between occupied and unoccupied subbands.
However, the exchange-correlation energies
in the fractional quantum Hall effect, for example, are only of the order of
$10^{-3}\hbar\omega_c$.
Thus linear response theory is unable to explain the cleanest experiments
in mesoscopics:  the quantization found in IQHE systems at large currents.

Let us now describe a general approach to nonlinear steady-state
mesoscopic transport which resolves this.
The thermodynamic variables which can be taken as known
are the internal energy $U$ and the particle number $N$.
We add to these the net current $I_m$ at each terminal
(so that $I_s=-I_d=I$ and $I_m=0, m\neq s,d$).
Following the MEA \cite{MEP},
we then maximize the information entropy
$S_I=-c\sum_\gamma p_{\gamma}\ln p_{\gamma}$,
subject to constraints on the average values of energy,
particle number, and currents.
Here $c$ is an (as yet) unspecified constant, and
$p_{\gamma}$ is the probability that the system is in a microstate $\gamma$.
This can be written as the matrix element
$p_{\gamma}=\langle \gamma | \hat\rho | \gamma \rangle$
of the density matrix $\hat\rho$.
Averages of an operator $\hat A$ are given by
$\langle \hat A \rangle \equiv
{\rm Tr}\left[ \hat A\hat \rho\right] /{\rm Tr} \hat \rho$.
The constraints are imposed by requiring that
$\langle \hat H \rangle = U$, $\langle \hat N \rangle = N$,
and $\langle {\hat I}_m \rangle = I_m$.  Here
$\hat H$, and $\hat N$ are, respectively, the Hamiltonian and
particle number operators, and ${\hat I}_m$ is the net current
operator in lead $m$ \cite{currentoperator}.
The constrained maximization gives the density matrix
\begin{equation}
\hat\rho=\exp{[-\beta(\hat H-\mu\hat N-\sum_m \xi_m {\hat
I_m})]}\label{rhohat}.
\end{equation}
In this expression $\mu$ is the {\em global} chemical potential,
associated with a global particle reservoir,
and the intensive variables $\xi_m$ are
Lagrangian multipliers associated with the constraints on the currents.
Because of current conservation there are only $M-1$
independent current constraints,
so we may choose $\xi_d=0$.
The variable $\beta$ is the product of $c^{-1}$ and the
variable conjugate to $U$.
This density matrix has the general form which Hershfield \cite{Hershfield}
recently showed exists quite generally in steady-state nonequilibrium systems;
following his work we therefore identify $\beta=1/k_BT$, where $k_B$ is
Boltzmann's constant and $T$ the thermodynamic temperature.
This identification also means that in this case the
information and thermodynamic entropies are identical (with $c=k_B$).

This formal result can be more clearly written in terms of
a complete set of single-particle eigenstates
$|\psi_\alpha\rangle$ of $\hat H$ and $\hat I_m$.
The above density operator then gives the following
thermal occupancies of single-particle states:
\begin{equation}
f_{\alpha}={1\over1+\exp{[\beta(\epsilon_{\alpha}-\mu-\sum_m\xi_m
i_{m,\alpha})]}}.\label{occ}
\end{equation}
For illustration, consider this result in terms of
the scattering states of a two-terminal
system. In this case we can drop the
terminal index $m$, and understand that $k>0$ corresponds to states
injected by the source and $k<0$ to states injected by the drain.
In an ideal system with $t_{n'k',nk}=\delta_{n'n}\delta_{k'k}$,
these states carry currents
$i_{nk}$, and
$f_{nk}=1/\left[e^{\beta(\epsilon_{nk}-\mu-\xi i_{nk})}+1\right]$,
where $\xi=\xi_s$.
In the simplest case, with only one subband ($n=0$)
occupied, this is similar to the LB result;
the combination $\mu+\xi i_{0k}t$
acts like an effective local chemical potential.
However, with more than one subband occupied, or with nontrivial $t$'s,
or at large voltages,
our current-constrained equilibrium occupations
cannot be described in terms of local chemical potentials.
In the general case states are occupied up to different
energies in each subband.

A voltmeter connected between the source and drain measures
the work required to move a unit charge
between them.  In the LB formalism this voltage is simply
$(\mu_{s}-\mu_{d})/e$.
Let us find the corresponding result for our distribution.
We use the representation given by the scattering states
$|\psi^+_{mnk}\rangle$, which have occupancies
$f_{mnk}$.
For clarity here we will present the results for a two-terminal device.
The generalization to the multi-terminal case is straightforward and
will be presented elsewhere.
The steady-state
condition and the absence of inelastic scattering within the device allow
us to define thermodynamic potentials of the electron distribution, just
as for an equilibrium system \cite{Tykodi}.  For example, the equivalence
in the present case between the
information and thermodynamic entropies
means that, as in equilibrium, here the Helmholtz free energy $F=U-TS$.
The thermodynamic work $\delta W$ done on the system at constant temperature
is then equal to the change in free energy,
$\delta{F} =\mu\,\delta N+\sum_m \xi_m\,\delta I_m$; in the
case of two terminals, this becomes $\delta{F} = \mu\,\delta N + \xi\,\delta
I$,
where $I$ is the source-to-drain current, and $\xi=\xi_s$.
Varying either of the variables $\mu$ and $\xi$
generally changes the occupancy of
states injected by
both terminals, by Eq.~(\ref{occ}).
Let $\eta$ refer to either of the variables $\mu,\xi$,
and let $\delta I^{\eta}$ be the change in net current
when $\eta$ is varied with the other variable held fixed.
Similarly let $\delta N^{\eta}_{m}$ be the change in the occupancy
of scattering states injected by terminal $m$ when $\eta$ is varied
(with $N_m=\sum_{nk}f_{mnk}$ the total particle number injected at
terminal $m$), so that
the change in free energy is
$\delta F^{\eta}= \mu (\delta N^{\eta}_s + \delta N^{\eta}_d)
               + \xi\,\delta I^{\eta}.
$
We obtain the potentials $V_{m}$ at the terminals by
interpreting this free energy change as the work done in adding electrons
$\delta N^\eta_m$
injected at each terminal against the voltage $V_m$ at the
terminal. Thus, a change $\delta N^\eta_m$ occurs at a cost in work
of $e\,\delta N^\eta_m V_m$.
The total work is then
$ e\,(\delta N^{\eta}_s V_s + \delta N^{\eta}_d V_d)$, and
equating this to $\delta F^{\eta}$ for
$\eta=\mu,\xi$ gives two linearly independent equations
\begin{equation}
\sum_{m=s,d} \delta N^{\eta}_{m} (eV_{m}-\mu)=
\xi\,\delta I^{\eta},
\label{lineq}
\end{equation}
which must be solved for the unknown terminal voltages $V_{s}$ and $V_d$.
(In the $M$-terminal
case, this becomes a set of $M$ equations.)
The resistance measured between source and drain is then
$R=(V_{s}-V_{d})/I$.
The $2\times2$ matrix $\delta N^{\eta}_{m}$ on the left-hand side of
Eq.~(\ref{lineq}) is
invertible, and the resulting potentials automatically are given relative to
the
global chemical potential $\mu$.

The distribution $f_{\alpha}$ in Eq.~(\ref{occ}) has been written
down in earlier work by Heinonen and Taylor \cite{HeinonenTaylor}, who
used it to study current distributions, and
more recently by Ng \cite{Ng}.  In these works it was argued that the
lack of dissipation in a device without inelastic scattering permitted
the ansatz of minimizing a free energy, subject to the current constraint.
Here we have shown how this can be justified much more generally within the
MEA, and the absence of dissipation
makes it possible to determine all microstates.
The second completely new point in the current work is our calculation
of voltage from considerations of work.  Ng, for example, simply assumed
that the current-induced potential difference is proportional
to the change in occupancies at the terminals.  The validity of this assumption
is not at all clear in, {\it e.g.}, precision IQHE measurements where the
Hall voltage is much greater than the bulk Fermi energy.
Furthermore, Ng
failed to constrain particle number, and consequently predicted that
even at small currents only states injected at the source (and not the
drain) should be occupied.  This appears unphysical.

We will illustrate our approach with two examples, turning first
to the resistance of an ideal two-terminal system.  For simplicity,  we
we drop the terminal subscripts, and
use eigenstates which satisfy periodic
boundary conditions on a length $L$
along the device.
(This is only to choose a simple density of states; the final result
does not depend on this particular boundary condition.)
Suppose that current-carrying states have energies
$\epsilon_{nk}=\epsilon_n + \hbar^2 k^2/2m^{*}$ and carry currents
$e\hbar k/ m^*L$.
This can represent 1D
transport, or a parabolically confined Hall bar.
The occupancies are,
by Eq.~(\ref{occ}), $f_{nk}=f(\epsilon_{nk}-\xi i_{nk}-\mu)$,
where $f(\epsilon)=1/(e^{\beta\epsilon}+1)$.
Then $f_{nk}$ is symmetric about $k=\tilde{\xi}\equiv \xi e/\hbar L$,
and we define
$\tilde{\epsilon}_{nk}=\epsilon_n+\hbar^2(k-\tilde{\xi})^2/2m^{*}$ so
that $\epsilon_{nk}-\mu-\xi i_{nk}=\widetilde\epsilon_{nk}-\widetilde\mu$,
where $\widetilde\mu=\mu+\hbar^2\tilde{\xi}^2/2m^{*}$. The electron number is
$N=\sum_n (2\pi/L)\int dk\, f(\widetilde\epsilon_{nk}-\widetilde\mu)$,
and the total current is
$I=\sum_n(2\pi/L)\int dk\, i_{nk}f(\widetilde\epsilon_{nk}-\widetilde\mu)$.
We convert
the integrals over $k$ to integrals over energy $\widetilde\epsilon_{nk}$,
and obtain
$N=\sum_n \int_{\epsilon_n}^{\infty} d{\widetilde\epsilon}
\rho_n({\widetilde\epsilon})
f({\widetilde\epsilon}-\tilde{\mu})$
where
$\rho_n(\tilde{\epsilon})=(L/\pi)[2\hbar^2(\tilde{\epsilon}-\epsilon_n)/m^{*}]^{-1/2}$ is the 1D density of states.
Similarly, $I=e\hbar/(m^*L)\widetilde\xi
\sum_{n}(2\pi/L)\int dk\, f(\widetilde\epsilon_{nk}-\widetilde\mu)$,
so $I=e\hbar N \tilde{\xi}/m^{*}L$.
It is then simple to calculate $\delta I^\eta$ and $\delta N^\eta_m$,
in  Eq.~(\ref{lineq}),  with the integrals for $\delta N^\eta_m$ over $k$
restricted to $k>0$ ($k<0$) for $m=s(d)$. The resulting expression for the
voltage difference $V_s-V_d$ obtained by inverting Eq.~(\ref{lineq}) is
simplified by the symmetry of $f_{nk}$ about $\widetilde\xi$, and we find
\begin{equation}
R = (V_s-V_d)/I = {h \over e^2}
\left(\sum_n {1\over e^{\beta(\epsilon_n-\mu)}+1}\right)^{-1}.
\label{R}
\end{equation}
If $\mu$ exceeds only the band minima of the first $j$
subbands (or Landau levels),
then at zero temperature $R=h/je^2$.  Finite-temperature corrections
are exponentially small.  This exact result is true {\it regardless
of the size of the voltage or current}.  This is perhaps our most
important result;  ours is the first mesoscopic transport theory which
can explain the extremely accurate quantization seen in the IQHE
far from the linear response regime.
(We neglect the breakdown which occurs in the IQHE at very large current
densities when other dissipative mechanisms turn on \cite{Cage}.)
If, for $I>0$, there are no states with $k<0$ occupied, $R$ is
not quantized; this appears to be the case with quantum point contacts
at large currents \cite{van Houten}.
We have numerically studied non-parabolic energies $\epsilon_{nk}$ and
multi-terminal systems
and find in these cases the accuracy of the quantization is limited
only by the numerical accuracy, so long as there are states with $k<0$
occupied.

As a second example, we consider a system with a `dirty' source
(a source with backscattering).
The LB formalism involves only Fermi surface properties
\cite{ADStone,BarangerStone}. As a consequence \cite{Buttiker3}, even in the
presence of such `dirty contacts', in the LB approach
all Hall conductances are quantized and
all longitudinal conductances are zero---provided that
no two dirty contacts are adjacent to one another.
This in fact can only be true in the linear response regime. At finite
current $I$, the net current at each terminal current involves an integral
over transmission probabilities, and resistances need then not be quantized.
Consider as a simple example
the four-terminal resistances in a system with a dirty source in the
presence of a magnetic field, with only the lowest
subband occupied.
We find that only at very small currents
do the resistances attain their ideal values (this is illustrated in
Fig.~\ref{dirty}).
Even the LB approach, if applied naively
beyond linear response, gives deviations in some resistances.
To inject a given current despite back-reflection at the source requires
$\mu_s$
to be greater than its value in the ideal case.  Then resistances between
the source and the other terminals differ from the ideal values.
Resistances not involving the source are still
ideal (quantized or zero) in the LB formalism.
In our approach, even the latter
are non-ideal at finite currents (see Fig.~\ref{dirty}) because the
occupancies of electrons
injected at terminal $m$ depend on the transmission from all
other terminals into $m$ [Eq.~(\ref{occ})].

In the MEA
observables enter the formalism as constraints.
Consequently, we have included the presence of a current $I$,
driven by a current source, as a constraint on the net current.
The ability of this to describe the IQHE at large currents is not
trivial and argues, we believe, for its validity.
We note that the LB formalism can also
be obtained from the MEA
if the current source is assumed to constrain the particle number
$N_m$
injected at
each terminal, rather than the current.  These constraints
are imposed by Lagrangian multipliers $\mu_m$, and the occupancies
which result are precisely the LB distributions
$f_{mnk}^{LB}$.
At low currents the use of local chemical potentials can be
justified using linear response theory, viewing the potential difference
(or the associated electric field) as driving the current.  This
cannot be extended to high currents.
Notice that in the MEA
the driving force need not be represented
by an operator in the Hamiltonian.  Instead, the {\it result} of the
driving (here, the current) enters as a constraint.
In the MEA the LB distribution would
arise if a current source could be thought of as an entity
that controls particle number instead of current.
One might suppose that
this models a voltage source instead of a current source.
If so, then the $I$-$V$ curve {\it at large currents and voltages}
would depend on whether voltage
or (as is usual) current is applied \cite{Ng}.
(In the linear regime, both approaches give the same result.)
This is possible in principle, since
voltage differences correspond to work, which is not a thermodynamic
state function. As we now show, this appears not to be the case, and
distributions of the form Eq. (\ref{occ}) should in fact be
expected in a steady-state dissipationless
system. Consider an arbitrarily long ideal device in which electrons
flow in their steady-state distributions with a net current.
In the device the Hamiltonian
(including electron-electron-electron interactions and
electron-phonon interactions) is translationally invariant
(neglecting Umklapp processes) and preserves the
distributions. In the frame of reference moving with velocity $v$
such that the
net current vanishes, the electrons are then in equilibrium at
some chemical potential $\mu'$. Hence the occupancy of a state with
energy $\epsilon_{nk}'$ is in this frame given by
$f({\epsilon'}_{nk}-\mu')$. By a Galilean transformation, $\epsilon_{nk}'
-\mu'=\epsilon_{nk}+\hbar kv-\mu=\epsilon_{nk}+\xi i_{nk}-\mu$,
where proportionality constants have been absorbed in $\xi$, and
$\mu$ differs from $\mu'$ by a constant. Since the occupancy
of a state is the same in each frame, the distributions in the
stationary frame
are thus $f(\epsilon_{nk}-\mu-\xi i_{nk})$.

The distributions Eq.~(\ref{occ}) lead to
other observable phenomena.  For example, they lead to dissipationless
deviations in quantization in the IQHE
when states in different subbands are mixed by short-range elastic
scatterers \cite{HJ}.  This might explain recent observations in
high-quality Si samples \cite{Yoshihiro}. We have also used our formalism
to explain the $I-V$ characteristics of quantum point contacts
\cite{van Houten} and
the lack of current saturation at high voltages \cite{QPC}.
The approach we have presented includes nonlinear effects due to the
current-dependent electron distributions.  At higher currents, other
nonlinearities arise from distortions of the electron wavefunctions by
the resulting electric field.  This field is due to electron-electron
interactions, which can easily be included in our approach at the Hartree
level.
In preliminary numerical calculations this causes
no qualitative change in the picture.

This work was supported in part by the UCF Division of Sponsored Research.

\begin{figure}\caption{The resistance $R_{sd,1d}$ measured between
terminal 1 and the drain when current flows from source to drain
in a four-terminal
IQHE system with a dirty source.
The device is a cross, with terminals 1 and 2 on opposite
sides of the current flow.  In an ideal system, this resistance
is $h/e^2$ at zero temperature.
The curves are at temperatures $T=0.0025$, 0.005, 0.01, 0.05, and 0.1
in units of $\hbar\omega_c/k_B$.
The zero-current Fermi energy
is $1.32$ $\hbar\omega_c$.
{\em Insert:\/} The
reflection probability $r$ at the source for the
lowest subband.}\label{dirty}\end{figure}

\begin{references}
\bibitem{convention} We use the convention that current into the device
is positive.
\bibitem{spinless} We neglect electron spins for simplicity.
\bibitem{currentoperator} $\hat I_m$ can be defined in terms of an integral
over matrix elements of the current density operator at terminal $m$, averaged
over the lead (see Ref. ~\onlinecite{Ng}).
The current operator so defined commutes with $\hat H$.
\end{references}
\end{document}